\documentclass[reprint,superscriptaddress,citeautoscript,floatfix]{revtex4-1}
\usepackage{xspace}

\usepackage{amsmath,amssymb,amsfonts,bbm,textcomp}
\usepackage[T1]{fontenc}
\usepackage{times}
\usepackage[shortcuts]{extdash}
\usepackage{graphicx}
\usepackage{hyperref}
\hypersetup{pdftitle=Zero-dimensional topologically nontrivial state in a superconducting quantum dot, pdfauthor=Pasquale Marra}
\usepackage[capitalize]{cleveref}
\usepackage{subfigure}

\newcommand{\pf}{\operatorname{pf}}
\newcommand{\pdet}{\operatorname{pdet}}

\newcommand{\sgn}{\operatorname{sgn}}
\newcommand{\up}{{\uparrow}}
\newcommand{\down}{{\downarrow}}

\newcommand{\nodag}{{\phantom{\dag}}}
\newcommand{\nod}{{\phantom{\dag}}}

\begin{document}
\title{%
Zero-dimensional topologically nontrivial state in a superconducting quantum dot
}
\author{Pasquale Marra}
\email{pasquale.marra@riken.jp}
\affiliation{RIKEN Center for Emergent Matter Science, Wakoshi, Saitama 351-0198, Japan}
\author{Alessandro Braggio}
\affiliation{NEST, Istituto Nanoscienze CNR and Scuola Normale Superiore, Piazza San Silvestro 12, 56127 Pisa, Italy}
\author{Roberta Citro}
\affiliation{Dipartimento di Fisica ``E. R. Caianiello'', Universit\`a di Salerno and CNR-SPIN, 84084 Fisciano (Salerno), Italy}

%\maketitle

\begin{abstract}
The classification of topological states of matter in terms of unitary symmetries and dimensionality predicts the existence of nontrivial topological states even in zero-dimensional systems, i.e., a system with a discrete energy spectrum.
Here, we show that a quantum dot coupled with two superconducting leads can realize a nontrivial zero-dimensional topological superconductor with broken time-reversal symmetry, which corresponds to the finite size limit of the one-dimensional topological superconductor.
Topological phase transitions corresponds to a change of the fermion parity, and to the presence of zero-energy modes and discontinuities in the current-phase relation at zero temperature.
These fermion parity transitions therefore can be revealed  by the current discontinuities or by a measure of the critical current at low temperatures.
\end{abstract}

\date{\today}

%\keywords{Josephson effect, Josephson junctions, superconducting quantum dots, quantum dots, topological states, topological superconductors}

\maketitle

\section{Introduction}

Since the discovery of the quantum Hall effect\cite{Klitzing1980,Thouless1982} and the theoretical prediction of Majorana bound states in triplet superconductors\cite{Kitaev2001}, a whole new class of novel electronic phases has been theoretically described and experimentally realized, namely, the class of topologically nontrivial states of matter\cite{Hasan2010,Qi2011,Wehling2014,Chiu2016}. 
Topological states of matter can be classified in terms of the antiunitary symmetries and dimensionality of the Hamiltonian\cite{Altland1997,Schnyder2009,Kitaev2009,Chiu2016}.
Analogously to the periodic table of chemical elements  in chemistry,  this classification has been a general guide to the discovery of novel topological phases  in solid state physics. 
Moreover, it predicts the existence of nontrivial topological states even in zero-dimensions, i.e. in a system with discrete energy spectrum.

A very important class of topological states of matter are topological superconductors:
These materials support Majorana zero-energy modes at the edges of the system\cite{Mourik2012,Nadj-Perge2014,Pawlak2016}, which have been proposed as the building block of topological quantum devices\cite{Alicea2012,Aguado2017,Lutchyn2018,sato_topological_2017,leijnse_introduction_2012,elliott_colloquium_2015,Beenakker2013Review}.
The simplest realization of a topological superconductor is the well-known Kitaev chain\cite{Kitaev2001} which can be implemented in a one-dimensional system proximized by a conventional superconductor in the presence of magnetic field and spin-orbit coupling\cite{Lutchyn2010,Oreg2010,Gangadharaiah2011,Klinovaja2012,Klinovaja2013}.
Moreover, topological superconductors exhibit very distinct features in their transport properties and in particular in their Josephson current\cite{Jiang2011,SanJose2012,Rokhinson2012,Brunetti2013,Chang2013,SanJose2013,Dolcini2015,Khanna2016,Peng2016,Hussein2016,Mellars2016,Wiedenmann2016,Jellinggaard2016,Zyuzin2016,Deacon2017,Virtanen2017,Mukherjee2017,Khanna2017,Pientka2017,Hyang2017,Lee2017,Cayao2017,Dominguez2017,Cayao2018}.

In a recent work\cite{Marra2016}, we have studied the short-size limit of a one-dimensional (1D) topological superconductor with broken time-reversal and chiral symmetries.
In this limit, the system turns zero-dimensional (0D), i.e., its energy spectrum is a finite set of discrete energy levels. 
This 0D superconductor exhibits topological phase transitions which correspond to variations of the fermion parity and to the occurrence of zero-energy modes which are a linear combination of particle and hole states\cite{Marra2016}. 
These fermion parity transitions can be revealed by discontinuities in the Josephson current-phase relation (CPR) in the zero-temperature limit.

\begin{figure}
\centering
\includegraphics[width=\columnwidth]{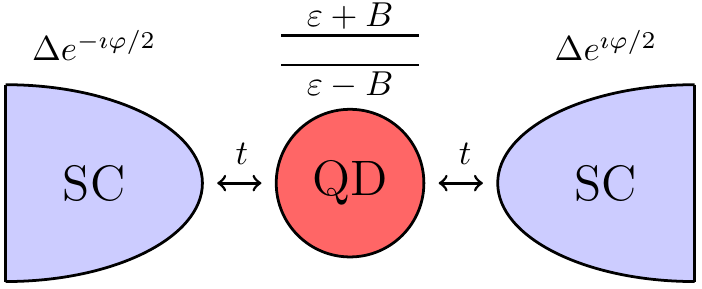}%
\\[-0.5mm]
\caption{
An SC-QD-SC Josephson junction realized by a two-level quantum dot in a magnetic field $B$ and electric gate $\varepsilon$ coupled with two superconducting leads.
The two energy levels are respectively $\varepsilon\pm B$.
The dot is coupled to the superconducting leads via tunneling junctions with transparency $t$.
The Josephson current $I_\varphi$ through the dot depends on the gauge-invariant phase difference $\varphi$ between the two superconducting leads.
}
\label{fig:System}
\end{figure}

Here we describe the simplest realization of such a 0D topological superconductor, i.e., a quantum dot\cite{Loss1998,Choi2000,Kouwenhoven2001,Wiel2001} coupled with two superconducting leads in a magnetic Zeeman field, forming a superconductor-quantum dot-superconductor (SC-QD-SC) Josephson junction. 
Zero-energy modes and the corresponding CPR discontinuities and ground-state parity crossings\cite{Tarasinski2015,Beenakker2013,Stanescu2013,Lee2013,Yokoyama2013,Yokoyama2014,Klees2017} have been recognized as precursors of Majorana modes in the long-wire limit\cite{Marra2016,SanJose2012}, and of Floquet Majorana modes realized in driven quantum dots\cite{Yantao2014,Benito2015}.
We will analytically derive and discuss the spectrum and the Josephson current of the dot, which agrees with the universal  prediction for zero-dimensional systems described in our previous work~\onlinecite{Marra2016}.
This allows us to reinterpret in terms of topological states the different regimes of the dot, which are already discussed in the literature\cite{Meng2009,Wentzell2016,Braggio2011,Droste2012,Hussein2016,Hussein2017}.
We will analyze in detail the relation between the topological properties of the groundstate, the zero-energy modes, and the corresponding CPR discontinuities.
We will show that, in this system, a topologically nontrivial state can be induced by a finite Zeeman field which breaks the time-reversal symmetry, even without a finite spin-orbit coupling. 
The resulting topological transitions coincide with a change of the fermion parity (topological invariant) and can be identified by discontinuities in the CPR and by a measure of the critical current at low temperatures. 

%\section{Results and discussions}
%\subsection{Effective model}
\section{Effective model}

We consider a semiconducting quantum dot in a magnetic field $B$ and coupled with two superconducting leads, as shown in \cref{fig:System}.
We assume that the only effect of the magnetic field is the lifting of the spin degeneracy via the Zeeman effect, and we neglect orbital effects of the  field.
Moreover we assume that the level spacing of the dot is larger than the Zeeman energy $B$ and than the Coulomb interaction $U$ within the dot, and therefore we neglect the contribution of higher energy levels.
Therefore we take into account only the levels $\varepsilon\pm B$ of the Kramers doublet closest to the Fermi energy. 
Here, $\varepsilon$ is the energy level of the dot in absence of Zeeman field, which can be modified by controlling the gate voltage.
This system can be described by a superconducting Anderson impurity model 
\begin{equation}
H=H_{\mathrm{QD}}+\sum_{i=L,R} H_i+H_{t_i}
\end{equation}
where the dot Hamiltonian is given by
\begin{gather}
H_\mathrm{QD}=
\varepsilon\, 
\begin{bmatrix} d_\up^\dag & d_\down^\dag \end{bmatrix}\cdot \begin{bmatrix} d_\up \\ d_\down \end{bmatrix}
+
B\, 
\begin{bmatrix} d_\up^\dag & d_\down^\dag \end{bmatrix}
\cdot  
\sigma_z 
\cdot
\begin{bmatrix} d_\up \\ d_\down \end{bmatrix}
+\nonumber\\+
U \left(n_\up-\frac12\right)  \left(n_\down-\frac12\right)
\label{eq:HamiltonianDot}
\end{gather}
where $d_\up^\dag,d_\down^\dag$ and $d_\up,d_\down$ are the creation and annihilation operators of the electrons in the dot, 
$n_\up=d_\up^\dag d_\up^\nod$ and $n_\down=d_\down^\dag d_\down^\nod$ the number operators, $\varepsilon\pm B$ the two-energy levels of the dot, and $U$ the onsite Coulomb repulsion. 
We assume hereafter that $e=\hbar=1$.

The Hamiltonians of the two superconducting leads $i=L,R$ are given by
\begin{gather}
H_i=
\sum_{\mathbf{k}}\varepsilon_{\mathbf{k}}\, 
 \begin{bmatrix}c_{\mathbf{k},i, \up}^\dag &c_{\mathbf{k},i, \down}^\dag \end{bmatrix}
\cdot
 \begin{bmatrix}
 c_{\mathbf{k},i,\up} \\ 
 c_{\mathbf{k},i,\down}
 \end{bmatrix}
+\nonumber\\+
\left(
\frac12\Delta e^{\imath\varphi_i}
 \begin{bmatrix}c_{\mathbf{k},i, \up}^\dag &c_{\mathbf{k},i, \down}^\dag \end{bmatrix}
 \cdot\left(
\imath\sigma_y \right)
\cdot
 \begin{bmatrix}c_{-\mathbf{k},i, \up}^\dag \\ c_{-\mathbf{k},i, \down}^\dag \end{bmatrix}
  + \mathrm{h.c.}
\right)
,
\label{eq:HamiltonianLead}
\end{gather}
where $c_{\mathbf{k},i, \up}^\dag, c_{\mathbf{k},i, \down}^\dag$ and $c_{\mathbf{k},i,\up}, c_{\mathbf{k},i,\down}$ are the creation and annihilation operators of electrons in the superconducting lead $i=L,R$ and with momentum $\mathbf{k}$,
 $\varepsilon_{\mathbf{k}}$ is the bare electron dispersion with respect to the Fermi level $\varepsilon_F=0$, 
$\Delta$ the magnitude of the superconducting gap, 
and $\varphi_i$ the phase of the superconducting gap in the two leads respectively. 
Here we assumed a standard BCS $s$-wave pairing and the same bare electron dispersion in the two superconducting leads.
In the following we furthermore assume that the bare electron dispersion varies in the interval $[-D,D]$ and that the density of states is  $\rho_0=1/(2D)$ with $2D$ the total bandwidth.

The tunneling between the dot and the leads is described by the tunnel Hamiltonians which read
\begin{equation}
H_{t_i}=
t \sum_\mathbf{k} 
 \begin{bmatrix}c_{\mathbf{k},i, \up}^\dag & c_{\mathbf{k},i, \down}^\dag \end{bmatrix}
\cdot
\begin{bmatrix} d_\up \\ d_\down \end{bmatrix}
+ \text{h.c.},
\label{eq:HamiltonianTunnel}
\end{equation}
where $t=t_L=t_R$ is the transparency of the dot-lead tunneling. 
We assume that the junction is symmetric and that the tunneling amplitudes do not depend on the electron momenta (wide band limit approximation). 

\begin{figure*}
\centering
\includegraphics[width=\textwidth]{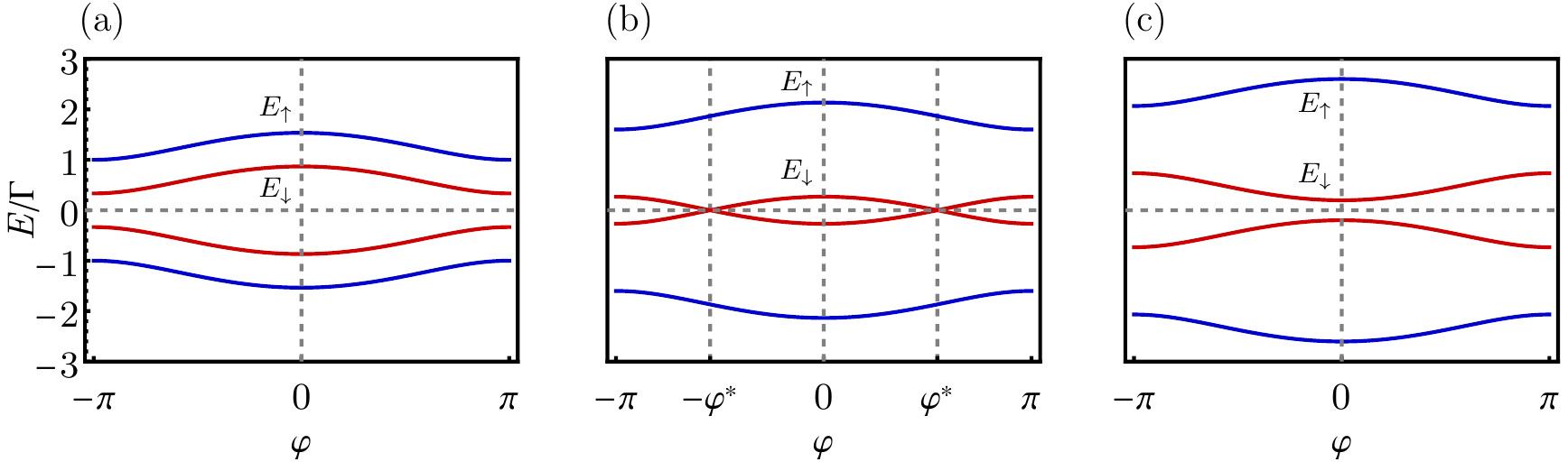}
\caption{
Energy spectrum of a two level quantum dot  coupled with two superconducting leads (SC-QD-SC junction), consisting of a set of four Andreev levels, i.e., two single-particle levels  $\pm E_{\up}$ (blue curves) and $\pm E_{\down}$ (red curves), as a function of the gauge-invariant phase difference $\varphi$ between the two superconducting leads.
We take $\varepsilon=2\Gamma/3$ and $U=0$.
The three panels corresponds to different values of the Zeeman field:
(a) small fields $|B|<B_\mathrm{min}$,
(b) intermediate fields $B_\mathrm{min}<|B|<B_\mathrm{max}$,
with the particle-hole gap closing at the gapless points $\pm\varphi^*$ [see \cref{eq:Gapless}],
and
(c) large fields $|B|>B_\mathrm{max}$.
}
\label{fig:Energy}
\end{figure*}

In the limit of a large superconducting gap, i.e., when the gap is larger than the characteristic frequencies of the quantum dot, the degrees of freedom of the leads can be effectively integrated  out\cite{Meng2009,Wentzell2016,Braggio2011,Droste2012,Hussein2016,Hussein2017}.
In absence of interactions ($U=0$)
 the system can be described by an effective Hamiltonian which reads\cite{Meng2009,Droste2012,Wentzell2016,Hussein2016,Hussein2017}
\begin{gather}
H_\mathrm{eff}=
\begin{bmatrix} d_\up^\dag & d_\down^\dag \end{bmatrix}
\cdot
\left(\varepsilon+B \sigma_z\right)
\cdot
\begin{bmatrix} d_\up \\ d_\down \end{bmatrix}
+\nonumber\\
+\Gamma\cos{(\varphi/2)}\,  
\begin{bmatrix} d_\up^\dag & d_\down^\dag \end{bmatrix}
\cdot \left(\imath\sigma_y\right) \cdot
\begin{bmatrix} d_\up^\dag \\ d_\down^\dag \end{bmatrix}
+ \text{h.c.},
\label{eq:HamiltonianEffective}
\end{gather}
where $\varphi=\varphi_R-\varphi_L$ is the gauge-invariant phase difference between the two leads, and where
\begin{equation}
\Gamma=4t^2\rho_0\arctan\left(\frac{D}\Delta\right)
\end{equation}
is the effective local superconducting pairing induced by the leads on the dot\cite{Meng2009,Wentzell2016}.
The Hamiltonian~\eqref{eq:HamiltonianEffective} can be written in the Bogoliubov-de Gennes formalism as
\begin{gather}
H_\mathrm{eff}=\Psi^\dag
\cdot
\begin{bmatrix}
\varepsilon+B\sigma_z&\Gamma\cos{(\varphi/2)}\,\imath\sigma_y\\
-\Gamma\cos{(\varphi/2)}\,\imath\sigma_y&-\varepsilon-B\sigma_z\\
\end{bmatrix}
\cdot
\Psi
\label{eq:BdGHamiltonianU0}
\end{gather}
where $\Psi^\dag=[d_\up^\dag,d_\down^\dag,d_\up^\nodag,d_\down^\nodag]$ and $\Psi=[d_\up^\nod,d_\down^\nod,d_\up^\dag,d_\down^\dag]^\intercal$ are the Nambu spinors describing the electron-hole pairs in the dot.
Notice that our definition of Nambu spinor differs from, e.g., Ref.~\onlinecite{Meng2009,Wentzell2016}, but it will allow us to define the topological invariant using the same formalism used in 1D superconductors.

The spectrum of this effective Hamiltonian is a set of four single particle states corresponding to two pairs of particle-hole symmetric Andreev levels $\pm E_\up$ and $\pm E_\down$ with
%\begin{subequations}
\begin{align}
E_\up&=E_\varphi+ B
\\\
E_\down&=E_\varphi- B
\quad\text{ with } E_\varphi=\sqrt{\varepsilon^2 + \Gamma^2 \cos^2(\varphi/2)},
\end{align}
%\end{subequations}
which correspond to the eigenstates described by the operators $\bar{d}^\dag_{\up\down}$
 defined by the Bogoliubov transformation
%\begin{subequations}
\begin{align}
\bar{d}^\dag_\up&=u d^\dag_\up + v d^\nod_\down,
\\
\bar{d}^\dag_\down&=u d^\dag_\down - v d^\nod_\up,
\end{align}
%\end{subequations}
where
%\begin{subequations}
\begin{align}
u&=\sqrt{(1+{\varepsilon}/{E_\varphi})/2},\\
v&=\sqrt{(1-{\varepsilon}/{E_\varphi})/2},
\end{align}
%\end{subequations}
The Bogoliubov factors  satisfy the properties $u^2+v^2=1$, $u^2-v^2=\varepsilon/E_\varphi$, and $uv= \Gamma |\cos(\varphi/2)|/(2E_\varphi)$.

Now we generalize the Hamiltonian \eqref{eq:BdGHamiltonianU0} to the case of finite interaction $U>0$.
A tedious but elementary calculation gives
$
(n_\up-1/2) (n_\down-1/2) =(\bar{n}_\up-1/2)  (\bar{n}_\down-1/2)
$
where  $\bar{n}_\up=\bar{d}_\up^\dag \bar{d}_\up^\nod$ and $\bar{n}_\down=\bar{d}_\down^\dag \bar{d}_\down^\nod$ are the number operators corresponding to the eigenstates of the effective Hamiltonian.
Therefore the Hamiltonian  in the presence of Coulomb interaction $U>0$ can be written in diagonal form as
\begin{equation}
\bar{H}_\mathrm{eff}=
\left(E_\varphi -\frac{U}2\right)(\bar{n}_\up+\bar{n}_\down)
+
B (\bar{n}_\up-\bar{n}_\down)
+
U \bar{n}_\up \bar{n}_\down,
\label{eq:HamiltonianEffectiveDiagonal}
\end{equation}
up to a numerical phase-independent constant.

The Hamiltonian eigenstates comprise the vacuum $|{00}\rangle$, the two single-particle states $|{01}\rangle$ and $|{10}\rangle$, and the two-particle state $|{11}\rangle$ with energies
%\begin{subequations}
\begin{align}
\bar{E}_{0}&=0,
\\
\bar{E}_{\down}&=E_\varphi-U/2-B,
\\
\bar{E}_{\up}&=E_\varphi-U/2+B,
\\
\bar{E}_{\up\down}&=2E_\varphi.
\end{align}
%\end{subequations}
Each of these particle states corresponds to a hole state by particle-hole symmetry.
The groundstate energy of the superconducting condensate is given by the sum of the single-particle energy 
levels\cite{Nazarov2009Page106}, which yield in this case
\begin{equation}
E_\mathrm{GS}(\varphi)=
|E_\varphi-U/2-B|+|E_\varphi-U/2+B|,
\label{eq:GSenergy}
\end{equation}
whereas the Josephson current at zero temperature is defined as $I_\varphi=-\partial_\varphi E_\mathrm{GS}(\varphi)$.
Notice that for small couplings $U/2<|\varepsilon|,|\Gamma|$, the only effect of the interaction is to shift the energy of the single-particle levels.
For this reason, if the conductance from the dot to the superconductor is relatively large (high dot-lead transparency) and one can consider the effect of interactions as a small perturbation.
Therefore the groundstate properties, such as the topological invariant and the Josephson current at zero temperature, are not affected in the case where $U/2<\varepsilon$ and $U/2<\Gamma$, as long as the particle-hole gap remains open and the Andreev levels do not cross.

In absence of interactions $U=0$,
the only possible groundstates are those with energies
\begin{equation}
E_\mathrm{GS}(\varphi)=
\begin{cases}
2E_\varphi &\text{\quad for \,} E_\up E_\down>0 ,
\\
2B &\text{\quad for \,} E_\up E_\down<0,
\end{cases}
\label{eq:GSenergyU0}
\end{equation}
which correspond respectively to the cases where the two single-particle levels $E_\up$ and $E_\down$ have the same sign or opposite sign.
We will show that the groundstate with energy $2E_\varphi$ is topologically trivial and has a finite Josephson current,
whereas the groundstate with energy $2B$ is topologically nontrivial and has a Josephson current which vanishes at zero temperature.

The phase diagram of this system has been already discussed in the literature\cite{Meng2009,Wentzell2016,Braggio2011,Droste2012,Hussein2016,Hussein2017}.
Since we consider here only the weak interacting case, we will not discuss the $0-\pi$ transition driven by the presence of  strong interaction.
A more thorough discussion of the the role of interactions on the 0D topological transition and on the ensuing  $\pi$-phase will be addressed in a following research paper.
Therefore, we will  discuss hereafter only quantum phase-transition in the regime of weak interactions in systems which can be described by \cref{eq:BdGHamiltonianU0} or \cref{eq:HamiltonianEffectiveDiagonal} for $U=0$.
Our findings cannot be applied to  $0-\pi$ transitions and to other kind of quantum phase-transitions which may be eventually present in this system, beyond the topological one we discussed.

%\subsection{The particle-hole gap and gapless points}
\section{The particle-hole gap and gapless points}

The particle-hole gap, i.e., the difference between the particle and hole levels closest to the Fermi level, closes if $|B|=E_\varphi$.
If one defines the two threshold fields $B_\mathrm{min}=|\varepsilon|$ and $B_\mathrm{max}=\sqrt{\varepsilon^2+\Gamma^2}$, one can verify that the spectrum is gapped for both small  $|B|<B_\mathrm{min}$ and large $|B|>B_\mathrm{max}$ Zeeman fields.
For intermediate fields $B_\mathrm{min}<|B|<B_\mathrm{max}$, the energy gap closes 
at specific values of the gauge invariant phase $\varphi=\pm\varphi^*$ where
\begin{gather}
\varphi^*=\arccos{(-\lambda)}
\text{\quad with }
\lambda=1+\frac{2(\varepsilon^2-B^2)}{\Gamma^2},
\label{eq:Gapless}
\end{gather}
where $|\lambda|<1$  
if $B_\mathrm{min}<|B|<B_\mathrm{max}$. 
We will show that these gapless points define a topological phase transition in the system, which corresponds to the appearance of discontinuous drops in the CPR of the junction. 

\Cref{fig:Energy} shows the
single-particle
energy spectrum of the system, i.e., the four particle-hole symmetric Andreev levels $\pm E_{\down}$ and $\pm E_{\up}$, as a function of the gauge-invariant phase difference $\varphi$.
As one can see, the energy spectrum is gapped for for small $|B|<B_\mathrm{min}$ and large $|B|>B_\mathrm{max}$ Zeeman fields respectively 
---  independently from the phase difference $\varphi$.
At intermediate fields $B_\mathrm{min}<|B|<B_\mathrm{max}$ the particle-hole gap closes at the gapless points $\pm\varphi^*$ which satisfy \cref{eq:Gapless}.
One can verify that the effect of a small Coulomb interaction $U/2<|\varepsilon|,|\Gamma|$ is a shift of the threshold fields $B_\mathrm{min}$ and $B_\mathrm{max}$ and of the value of the phases $\pm\varphi^*$ where the gap closes.

%\subsection{Topological invariant}
\section{Topological invariant}

\begin{figure}
\centering
\includegraphics[width=\columnwidth]{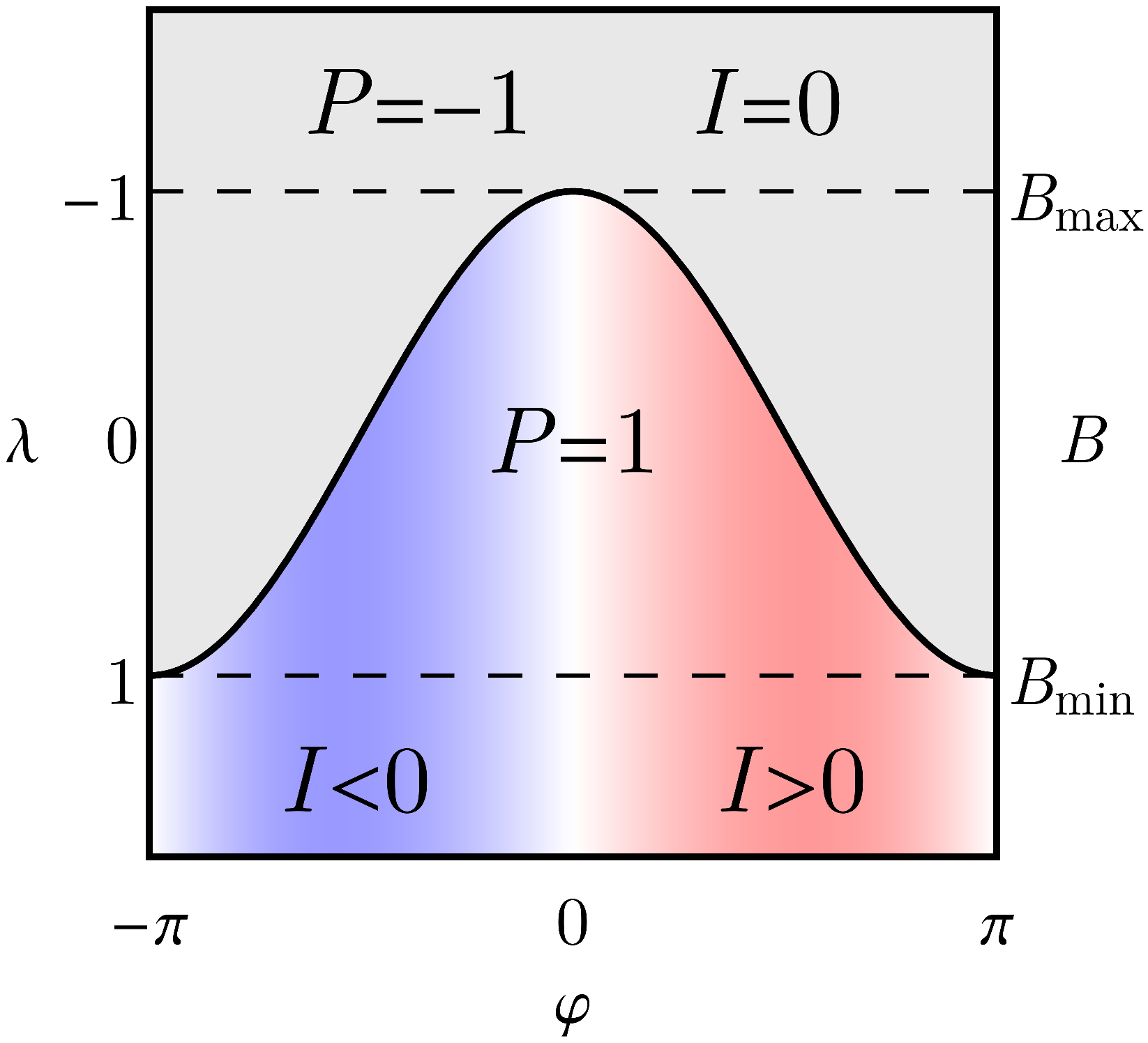}
\caption{
Topological phase space of a 0D topological superconductor realized by a quantum dot  coupled with two superconducting leads (SC-QD-SC junction).
The system realizes respectively a trivial state $P=1$ for small Zeeman fields $|B|<B_\mathrm{min}$ (i.e., $\lambda>1$), and a nontrivial state $P=-1$ for large fields $|B|>B_\mathrm{max}$ (i.e., $\lambda<-1$).
Notice that the Josephson current vanishes in the nontrivial state. 
Topological transitions coincides with the occurrence of 
zero-energy modes at $\pm\varphi^*=\pm\arccos(-\lambda)$ (solid line) 
for intermediate fields
$B_\mathrm{min}<|B|<B_\mathrm{max}$ 
(i.e., $|\lambda|<1$).
In this case the system is in its trivial $P=1$  and nontrivial $P=-1$  state respectively for $|\varphi|\lessgtr\varphi^*$ within the interval $\varphi\in[-\pi,\pi]$.
}
\label{fig:PhaseSpace}
\end{figure}

This simple 0D two-level system can realize a topologically nontrivial state which breaks time-reversal symmetry while preserving particle-hole symmetry. 
This topologically nontrivial state can be seen as
the 0D limit of a 1D topological superconductor, and as the minimal model for the system described in Ref.~\onlinecite{Marra2016}.
In fact, for finite Zeeman energies ($B\neq0$) and superconducting pairing ($\Gamma>0$), the system is in the Altland-Zirnbauer\cite{Altland1997,Schnyder2009,Kitaev2009,Chiu2016} symmetry class D (particle-hole symmetry, broken time-reversal and chiral symmetries). 
This class is characterized in 0D by a $\mathbb{Z}_2$ topological invariant which is defined in the non-interacting case $U=0$ as the fermion parity of the groundstate\cite{Loring2015,Marra2016} $P=\sgn\pf\left({H}_\mathrm{eff}\imath\tau_x\right)$, i.e., as the sign of the Pfaffian of the Hamiltonian in Majorana representation ($\tau_x$ is the first Pauli matrix in the particle-hole space).
The fermion parity labels the topological inequivalent groundstates as a function of the gauge-invariant phase $\varphi$, i.e., the trivial $P=1$ (even parity) and nontrivial state $P=-1$ (odd parity). 
The fermion parity of the 0D topological  quantum dot described by Hamiltonian~\eqref{eq:BdGHamiltonianU0} can be evaluated analytically. 
The square of the Pfaffian of a matrix is equal to the determinant, which is equal to the product of its eigenvalues, and therefore one has
$\pf(H_\mathrm{eff}\imath\tau_x)^2=\det(H_\mathrm{eff}\imath\tau_x)=\det(H_\mathrm{eff})=E_\up^2 E_\down^2$ due to particle-hole symmetry.
A direct calculation of the Pfaffian indeed shows that $\pf(H_\mathrm{eff}\imath\tau_x)=E_\up E_\down$ and therefore
\begin{equation}
P_\varphi
%=\sgn\pf\left[{H}_\mathrm{eff}\imath\tau_x\right]
=\sgn(E_\up E_\down)
=\sgn(E_\varphi^2-B^2)=
\sgn (\lambda+\cos\varphi)
,
\label{eq:FermionParity}
\end{equation}
where we used the definition of $\lambda$ given in \cref{eq:Gapless}.
This equation is a special case of Eq.~(2) of Ref.~\onlinecite{Marra2016}.
Notice that if $B=0$ the time-reversal symmetry is unbroken and the groundstate is trivial
$
P_\varphi
=\sgn(E_\varphi^2)=1
$
as expected.
As anticipated, the groundstate with energy $2E_\varphi$ is topologically trivial, since in this case $E_\up E_\down>0$,
whereas the groundstate with energy $2B$ is topologically nontrivial, since in this case one has $E_\up E_\down<0$.
Therefore, the inversion of the lowest-energy Andreev level corresponds to a topological transition to the nontrivial state.
The fermion parity defines the topological phase space of the system, and is completely determined by the gauge-invariant phase $\varphi$ and by the adimensional quantity $\lambda$, as shown in \cref{fig:PhaseSpace}.
Moreover, since $P=\sgn[E_\up E_\down]$, the condition $P_\varphi\equiv0$ corresponds to the gapless points $\varphi=\pm\varphi^*$ where zero-energy modes occur (solid line in \cref{fig:PhaseSpace}).

At small Zeeman fields $|B|<B_\mathrm{min}$ (i.e., $\lambda>1$), the system is in the topologically trivial state with even fermion parity $P=1$ for any value of the phase $\varphi$.
At large fields $|B|>B_\mathrm{max}$ instead (i.e., $\lambda<-1$), the system realizes the topologically nontrivial state with odd fermion parity $P=-1$ for any value of the phase $\varphi$. 
However, for intermediate $B_\mathrm{min}<|B|<B_\mathrm{max}$ (i.e., $|\lambda|<1$) topological transitions occur at the gapless points $\pm\varphi^*$ [see \cref{eq:Gapless}].
In this case the system realizes the trivial or in the nontrivial state (even or odd parity) respectively for $|\varphi|<\varphi^*$ and $|\varphi|>\varphi^*$ in the interval $\varphi\in[-\pi,\pi]$, as one can see in \cref{fig:PhaseSpace}.
The two gapless points $\pm\varphi^*$ therefore correspond to a quantum phase transition where the fermion parity of the groundstate changes from trivial to nontrivial.
Notice that for $|B|=B_\mathrm{min}$ and for $|B|=B_\mathrm{max}$ (i.e., $|\lambda|=1$)
no topological transition occurs, and the system is respectively in the trivial or nontrivial gapped state with the exceptions of the single gapless point $\varphi^*=\pi$ or $\varphi^*=0$  respectively.

Notice also that the particle-hole gap can close also in absence of Zeeman field if $\varepsilon=0$.
For $B=\varepsilon=0$ in fact (which gives $\lambda=1$) the gap closes at $\varphi^*=\pi$.
In this case the time-reversal symmetry is unbroken, and the system is gapped and topologically trivial for any value of the phase $\varphi\neq\pi$.

The topological phase space derived in the case of a superconducting quantum dot is universal for the class of zero-dimensional superconductors.
It coincide in fact with the topological phase space in Fig.~2(a) of Ref.~\onlinecite{Marra2016}, where it was derived in the more general case of a zero-dimensional quantum system (short-size regime) with an arbitrary number of energy modes.
Notice that the topological phases can be defined also in the case of small Coulomb interactions as long as the particle-hole gap remains open.
In this case in fact the topological invariant cannot change, since the phase with small interaction $U>0$
can be transformed with the non-interacting phase $U=0$ 
 by a smooth transformation without closing the gap.

It is important to note that in the 0D case (differently from the 1D case) topological states can be realized without spin-orbit coupling: 
This is because topological states in the symmetry class D are enforced by the presence of the superconducting coupling (particle-hole symmetry) and the Zeeman field (which breaks the time-reversal symmetry). 
The gap opening, in this case, is guaranteed in general by the gap induced by finite size effects or eventually by interactions.

%\subsection{Josephson current-phase discontinuities}
\section{Josephson current-phase discontinuities}

In our previous work\cite{Marra2016}, we have found the general relation between the topological invariant of a 0D topological superconductor and the discontinuities of the Josephson CPR.
The topological phase transition between the trivial ($P=1$, even fermion parity) and the nontrivial state ($P=-1$, odd fermion parity) corresponds to the emergence of a discontinuity in the Josephson CPR at zero temperature.
In this case in fact, the current is proportional to the phase-derivative of the total energy of the superconducting condensate\cite{Golubov2004,Nazarov2009Page106}, which is given by the sum of the positive energy levels $|E_\up|+|E_\down|$.
Hence, the Josephson current is equal to $-2\partial_\varphi E_\varphi$ in the trivial groundstate with  energy $E_\mathrm{GS}(\varphi)=2E_\varphi$,
whereas it vanishes in the nontrivial groundstate with energy $E_\mathrm{GS}(\varphi)=2B$ (see \cref{eq:GSenergyU0}).
The CPR at zero temperature is therefore given by
\begin{equation}
I_\varphi
=-[1+P_\varphi]\partial_\varphi E_\varphi
=[1+P_\varphi] \frac{\Gamma^2\sin\varphi}{4 E_\varphi}.
\label{eq:JosephsonT0}
\end{equation}
In the topologically trivial state ($P=1$) at low fields $|B|<B_\mathrm{min}$, the two energy levels $E_\up$ and $E_\down$ contribute equally to the Josephson current and one has $I_\varphi=-2\partial_\varphi E_\varphi$.
However, when the fermion parity changes, one of the energy level crosses the particle-hole gap, and its contribution to the current changes its sign.

Therefore, in the topologically nontrivial state ($P=-1$) at high fields $|B|>B_\mathrm{max}$ 
the Josephson current in \cref{eq:JosephsonT0} vanishes  since
the contributions from the two energy levels $E_\up$ and $E_\down$ cancel each other. 
Moreover, as one can see from \cref{eq:JosephsonT0}, 
for intermediate fields $B_\mathrm{min}<|B|<B_\mathrm{max}$,
(i.e., $|\lambda|<1$)
the CPR exhibits a discontinuity between the trivial state with $I=\pm2\Gamma^2\sin\varphi^*/[4E_{\varphi^*}]$ to the nontrivial one with $I=0$ at the gapless points $\pm\varphi^*$ which is equal to
\begin{equation}
\Delta I=
\frac{\Gamma^2\sqrt{1-\lambda^2}}{2|B|}
,
\label{eq:Discontinuity}
\end{equation}
which is a special case of Eq.~(3) of Ref.~\onlinecite{Marra2016}. 
The discontinuity is a consequence of the crossing at zero-energy of the lowest-energy level with linear phase-dispersion. 
The discontinuity in \cref{eq:Discontinuity} can be indeed also calculated directly using Eq.~(3) of Ref.~\onlinecite{Marra2016}, which can be rewritten as
\begin{equation}
\Delta I
=
2
\left.
\frac{|\partial_\varphi \pf(\mathcal{H}_\mathrm{eff}\imath\tau_x)|}
{\sqrt{|\pdet(\mathcal{H}_\mathrm{eff})|}}
\right\vert_{\varphi=\varphi^*}
,
\label{eq:DiscontinuityPRBR}
\end{equation}
where $\pdet(\mathcal{H}_\mathrm{eff})$ is the pseudodeterminant of the Hamiltonian (the product of nonzero eigenvalues). 
The square root of the pseudodeterminant is in this case just the product of the positive eigenvalues (due to particle hole-symmetry).
Since the system has only two non-negative single-particle energy levels $|E_\up|=|B+ E_{\varphi^*}|$ and $|E_\down|=|B- E_{\varphi^*}|$, and one of these two energy levels vanishes at gapless points $\pm\varphi^*$ since in this case $|B|=|E_{\varphi^*}|$, the denominator of \cref{eq:DiscontinuityPRBR} is equal to the nonzero positive energy level given by $|B|+|E_{\varphi^*}|=2|B|$, which yields ${\sqrt{|\pdet(\mathcal{H}_\mathrm{eff})|}}=2|B|$ which leads via \cref{eq:DiscontinuityPRBR} to \cref{eq:Discontinuity}.

\begin{figure*}
\centering
\includegraphics[width=\textwidth]{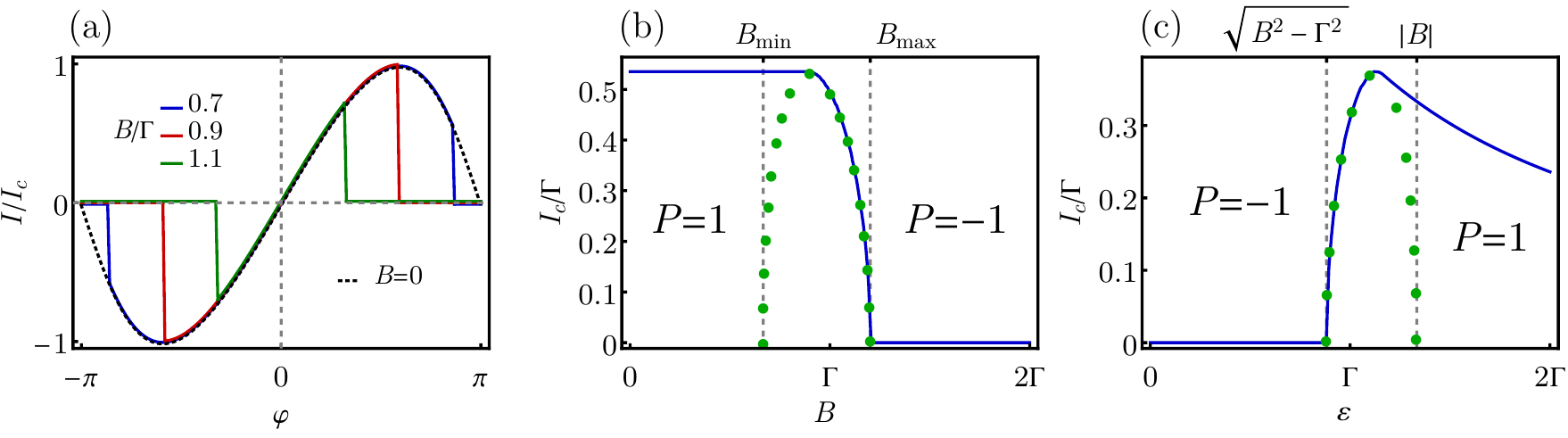}
\caption{
(a) Josephson CPR of the SC-QD-SC junction for different choices of the Zeeman field $B$ in the limit $T\to0$ [\cref{eq:JosephsonT0}] in units of the critical current of the trivial branch.
We take $\varepsilon=2\Gamma/3$.
Depending on the Zeeman field, different regimes are realized:
At small fields $|B|<B_\mathrm{min}$
(i.e, $\lambda>1$, dotted line)
the current is smoothly oscillating as a function of the phase $\varphi$ and the system is topologically trivial ($P=1$);
At large fields $|B|>B_\mathrm{max}$
(i.e., $\lambda<-1$, not shown)
the current vanishes and the system is topologically nontrivial ($P=-1$);
At intermediate fields $B_\mathrm{min}<|B|<B_\mathrm{max}$
(i.e., $|\lambda|<1$, solid lines), discontinuous drops appear at the transition points between the trivial and nontrivial topological states.
Current discontinuities correspond to the variations of the fermion parity and to the presence of zero energy modes. 
(b)
Critical current of the SC-QD-SC junction as a function of the Zeeman field at zero temperature (solid line) with $\varepsilon=2\Gamma/3$. 
(c)
Critical current of the SC-QD-SC junction as a function of the electric gate $\varepsilon$ at zero temperature (solid line) with $B=4\Gamma/3$. 
In both cases, 
the critical current drops from a finite value in the trivial state ($P=1$ and $\lambda>1$) to zero in the nontrivial state ($P=-1$ and $\lambda<-1$).
In the transition regions $B_\mathrm{min}<B<B_\mathrm{max}$ (b) 
and $\sqrt{B^2-\Gamma^2}<|\varepsilon|<|B|$ (c), the trivial and nontrivial states alternate at different phases $\varphi$.
As one can see,  
when the system approaches its nontrivial state $P=-1$,
the critical current coincides with the magnitude of the discontinuous drop $\Delta I$ (green dots) given in \cref{eq:Discontinuity}.
}
\label{fig:Josephson}
\end{figure*}

\Cref{fig:Josephson}(a) shows the CPR of the SC-QD-SC junction for different choices of the Zeeman field $B$ at zero temperature, calculated directly from \cref{eq:JosephsonT0}.
At low fields $|B|<B_\mathrm{min}$ (i.e., $\lambda>1$ ) the system is topologically trivial ($P=1$) and the CPR is smoothly oscillating without any discontinuity.
At large fields $|B|>B_\mathrm{max}$  (i.e., $\lambda<-1$), the system is topologically nontrivial ($P=-1$) and the Josephson current vanishes due to the opposite contribution of the two Andreev levels.
At intermediate fields $B_\mathrm{min}<|B|<B_\mathrm{max}$ instead (i.e., $|\lambda|<1$), discontinuities appear at the transition points between the trivial and nontrivial topological states (gapless points $\pm\varphi^*$).
The emergence of a discontinuous drop coincides with a change of the fermion parity and to the presence of zero-energy states closing the particle-hole gap.
Notice that, since the energy levels of the system depends smoothly on the phase $\varphi$, gapless points are the only points where the CPR can be discontinuous.
Notice that at finite temperatures, CPR discontinuities are smoothed out by the effect of thermal fluctuations.
However, such discontinuities can be revealed, e.g., by the presence of spikes in the phase-derivative of the CPR at low temperatures\cite{Marra2016}.

Hence, if time-reversal symmetry is broken ($B\neq0$), current discontinuities correspond to the presence of zero-energy modes and to a change in the topological invariant.
These signatures are topologically robust against small perturbations, such as disorder.
This means that these discontinuities and the associated zero-energy modes cannot be removed by the presence of, e.g., disorder or interactions, if these perturbations are small compared to the effective local pairing $\Gamma$ and Zeeman energy $B$.
The only effect of these small perturbations is in fact to produce a 
shift of the gapless point $\varphi^*\rightarrow \varphi^*+\delta\lambda/\sqrt{1-\lambda^2}$ where the topological transition and zero-energy modes occurs.
Notice also that discontinuities in the Josephson CPR  are still present in the interacting case\cite{Wentzell2016} at zero temperature.
As shown in Ref.~\onlinecite{Marra2016} in fact, the correspondence between CPR discontinuities and fermion parity transitions relies only on the presence of a broken time-reversal symmetry which removes the spin degeneracy and on the fact that in this case the closing of the particle-hole gap correspond to a change of the topological invariant.

On the other hand, if time-reversal symmetry is unbroken, current discontinuities are still present if $B=\varepsilon=0$ (where $\lambda=1$).
In this case, the CPR exhibits a single discontinuous drop $\Delta I=\Gamma/2$ at the gapless point $\varphi^*=\pi$, according to \cref{eq:Discontinuity}. 
This case reproduces the well-known current-phase discontinuity of a quantum point contact\cite{Golubov2004}.
However, in this case the discontinuity does not correspond to a topological transition.

Notice that the presence of a small Coulomb interaction does not affect the Josephson current at zero temperature in the trivial and non-trivial branches of the CPR, since the energy shift $U/2$ of the Andreev levels do not depend on the phase $\varphi$.

%\subsection{Critical current}
\section{Critical current}

The topological transition can be probed also by a measure of the critical current of the junction. 
The critical current is defined as the maximum current of the junction up to the phase $I_c=\max I_\varphi$.
In the trivial state at low fields $|B|<B_\mathrm{min}$ (i.e., $\lambda>1$) the critical current is finite.
Since the CPR is continuous in this case,
the maximum of the current coincides with the local maximum of the current where its phase-derivative vanishes $\partial_\varphi I_\varphi=0$.
In the limits $\varepsilon\to0$ and  $\varepsilon\to\pm\Gamma$ for example, 
the current reaches its maximum at $\widetilde{\varphi}=\pi$ or at $\widetilde{\varphi}=\pi/2$,
which gives a critical current $I_c=\Gamma/2$ and $I_c=\Gamma^2/(4\sqrt{\varepsilon^2+\Gamma^2/2})$ respectively.  
In the nontrivial state at large fields $|B|>B_\mathrm{max}$ instead ($\lambda<-1$) the current vanishes and one has $I_c=0$. 
However, at intermediate fields $B_\mathrm{min}<|B|<B_\mathrm{max}$ (i.e., $|\lambda|<1$) trivial and nontrivial states alternates in the interval $\varphi\in[-\pi,\pi]$, and the CPR has discontinuities.
In this case the CPR is not continuous, and therefore the maximum of the current may coincide either with the local maximum $I_{\widetilde{\varphi}}$ of the current where $\partial_\varphi I_\varphi=0$,
or with the current at the discontinuity $I_{\varphi^*}=\Delta I$.
More precisely, the critical current coincides with the maximum between these two values $I_c=\max(|I_{\widetilde{\varphi}}|,|\Delta I|)$.
The case $I_c=|\Delta I|$  occurs, for instance, when the system approaches its nontrivial state at large fields $|B|\to B_\mathrm{max}$.
Therefore for fields $|B|\lesssim B_\mathrm{max}$ the critical current coincide with the current discontinuity  $I_c=\Delta I$.
Notice that this regime can be obtained either by a measure of the critical current by varying the magnetic field, or by varying, e.g., the energy level $\varepsilon$ in a constant field $B$.

\Cref{fig:Josephson}(b) shows the critical current of the junction as a function of the Zeeman field. 
As one can see, the critical current is finite in the trivial $P=1$ state when $|B|<B_\mathrm{min}$ (i.e., $\lambda>1$), and drops to zero in the nontrivial $P=-1$ state when $|B|>B_\mathrm{max}$ (i.e., $\lambda<-1$) state.
The drop of the critical current is smooth in the intermediate region where $B_\mathrm{min}<|B|<B_\mathrm{max}$ (i.e., $|\lambda|<1$).
Analogously, \cref{fig:Josephson}(c) shows the critical current of the junction as a function of the electric gate $\varepsilon$ at at constant field $B$. 
The  smooth transition is obtained
 for intermediate values $\sqrt{B^2-\Gamma^2}<\varepsilon<|B|$ the Zeeman field varies in the range $B_\mathrm{min}<|B|<B_\mathrm{max}$, where we remind that $B_\mathrm{min}=|\varepsilon|$ and $B_\mathrm{max}=\sqrt{\varepsilon^2+\Gamma^2}$.
In the the intermediate region, when the system approaches its nontrivial state, the critical current coincide with the magnitude of the discontinuous drop $I_c=|\Delta I|$ (dots in the figures).
Hence, a measure of the critical current at low temperature can be used to indirectly probe the magnitude of the discontinuous drop and the existence of topological phase transitions and zero-energy modes even when a direct measure of the CPR is not accessible\cite{Tiira2017}.
Notice that it is reasonable to speculate that the current discontinuities may indicate a topological transition also in the interacting case.

\section{Conclusions}

We have shown that a quantum dot coupled with two superconducting leads can realize a 0D topological superconductor with broken time-reversal symmetry.
In this system, topological phase transitions between trivial and nontrivial states correspond to discontinuities in the Josephson CPR at low temperatures and to the presence of zero-energy modes.
This simple model, which can be treated analytically, fully confirms the results obtained in a more general model in Ref.~\onlinecite{Marra2016}.

The topological phase transitions and the ensuing current discontinuities are robust, in the sense that cannot be removed by small perturbations.
A direct measure of the CPR\cite{Golubov2004,Frolov2004,Sochnikov2013,Szombati2016} or  of the Josephson radiation\cite{Ozyuzer2007,Solinas2015,Deacon2017} at low temperatures can reveal the presence of such discontinuities. 
Moreover, the presence of the topological transition can be probed indirectly by a measure of the critical current of the junction as a function of the Zeeman field or gate voltage. 

\begin{acknowledgements}
P.~M. acknowledges financial support from JSPS Kakenhi Grant No. 16H02204.
A.~B. acknowledges the Italian MIUR-FIRB 2012 via the HybridNanoDev project under Grant no. RBFR1236VV, 
the European Research Council under the
European Union's Seventh Framework Program (FP7/ 2007-2013) ERC Grant No. 615187-COMANCHE for partial financial support, the CNR-CONICET cooperation programme ``Energy conversion in quantum nanoscale hybrid devices'', 
and the Royal Society though the International Exchanges between the UK and Italy (grant IES\ R3\ 170054).
\end{acknowledgements}

%\bibliographystyle{prsty_no_etal_titles_doi}
%\bibliography{bib}

\begin{thebibliography}{10}

\bibitem{Klitzing1980}
K.~v. Klitzing, G. Dorda, and M. Pepper, {\em New method for high-accuracy
  determination of the fine-structure constant based on quantized Hall
  resistance}, \href{http://dx.doi.org/10.1103/PhysRevLett.45.494}{Phys. Rev.
  Lett. {\bf 45},  494 } (1980).

\bibitem{Thouless1982}
D.~J. Thouless, M. Kohmoto, M.~P. Nightingale, and M. den Nijs, {\em Quantized
  Hall conductance in a two-dimensional periodic potential},
  \href{http://dx.doi.org/10.1103/PhysRevLett.49.405}{Phys. Rev. Lett. {\bf
  49},  405 } (1982).

\bibitem{Kitaev2001}
A. Kitaev, {\em Unpaired Majorana fermions in quantum wires},
  \href{http://dx.doi.org/10.1070/1063-7869/44/10S/S29}{Phys. Usp. {\bf 44},
  131 } (2001).

\bibitem{Hasan2010}
M.~Z. Hasan and C.~L. Kane, {\em \textit{Colloquium}: Topological insulators},
  \href{http://dx.doi.org/10.1103/RevModPhys.82.3045}{Rev. Mod. Phys. {\bf 82},
   3045 } (2010).

\bibitem{Qi2011}
X.-L. Qi and S.-C. Zhang, {\em Topological insulators and superconductors},
  \href{http://dx.doi.org/10.1103/RevModPhys.83.1057}{Rev. Mod. Phys. {\bf 83},
   1057 } (2011).

\bibitem{Wehling2014}
T.~O. Wehling, A.~M. Black-Schaffer, and A.~V. Balatsky, {\em Dirac materials},
  \href{http://dx.doi.org/10.1080/00018732.2014.927109}{Adv. Phys. {\bf 63},  1
  } (2014).

\bibitem{Chiu2016}
C.-K. Chiu, J.~C.~Y. Teo, A.~P. Schnyder, and S. Ryu, {\em Classification of
  topological quantum matter with symmetries},
  \href{http://dx.doi.org/10.1103/RevModPhys.88.035005}{Rev. Mod. Phys. {\bf
  88},  035005 } (2016).

\bibitem{Altland1997}
A. Altland and M.~R. Zirnbauer, {\em Nonstandard symmetry classes in mesoscopic
  normal\-/superconducting hybrid structures},
  \href{http://dx.doi.org/10.1103/PhysRevB.55.1142}{Phys. Rev. B {\bf 55},
  1142 } (1997).

\bibitem{Schnyder2009}
A.~P. Schnyder, S. Ryu, A. Furusaki, and A.~W.~W. Ludwig, {\em Classification
  of topological insulators and superconductors},
  \href{http://dx.doi.org/10.1063/1.3149481}{AIP Conf. Proc. {\bf 1134},  10 }
  (2009).

\bibitem{Kitaev2009}
A. Kitaev, {\em Periodic table for topological insulators and superconductors},
  \href{http://dx.doi.org/10.1063/1.3149495}{AIP Conf. Proc. {\bf 1134},  22 }
  (2009).

\bibitem{Mourik2012}
V. Mourik, K. Zuo, S.~M. Frolov, S.~R. Plissard, E.~P. A.~M. Bakkers, and L.~P.
  Kouwenhoven, {\em Signatures of Majorana fermions in hybrid
  superconductor\-/semiconductor nanowire devices},
  \href{http://dx.doi.org/10.1126/science.1222360}{Science {\bf 336},  1003 }
  (2012).

\bibitem{Nadj-Perge2014}
S. Nadj-Perge, I.~K. Drozdov, J. Li, H. Chen, S. Jeon, J. Seo, A.~H. MacDonald,
  B.~A. Bernevig, and A. Yazdani, {\em Observation of Majorana fermions in
  ferromagnetic atomic chains on a superconductor},
  \href{http://dx.doi.org/10.1126/science.1259327}{Science {\bf 346},  602 }
  (2014).

\bibitem{Pawlak2016}
R. Pawlak, M. Kisiel, J. Klinovaja, T. Meier, S. Kawai, T. Glatzel, D. Loss,
  and E. Meyer, {\em Probing atomic structure and Majorana wavefunctions in
  mono-atomic Fe chains on superconducting Pb surface}, Npj Quantum Information
   {\bf 2},  16035 EP   (2016).

\bibitem{Alicea2012}
J. Alicea, {\em New directions in the pursuit of Majorana fermions in solid
  state systems}, \href{http://dx.doi.org/10.1088/0034-4885/75/7/076501}{Rep.
  Prog. Phys. {\bf 75},  076501 } (2012).

\bibitem{Aguado2017}
R. Aguado, {\em Majorana quasiparticles in condensed matter},
  \href{http://dx.doi.org/10.1393/ncr/i2017-10141-9}{La Rivista del Nuovo
  Cimento {\bf 40},  523 } (2017).

\bibitem{Lutchyn2018}
R.~M. Lutchyn, E.~P. A.~M. Bakkers, L.~P. Kouwenhoven, P. Krogstrup, C.~M.
  Marcus, and Y. Oreg, {\em Majorana zero modes in
  superconductor--semiconductor heterostructures},
  \href{http://dx.doi.org/10.1038/s41578-018-0003-1}{Nature Reviews Materials
  {\bf 3},  52 } (2018).

\bibitem{sato_topological_2017}
M. Sato and Y. Ando, {\em Topological superconductors: a review},
  \href{http://dx.doi.org/10.1088/1361-6633/aa6ac7}{Rep. Prog. Phys. {\bf 80},
  076501 } (2017).

\bibitem{leijnse_introduction_2012}
M. Leijnse and K. Flensberg, {\em Introduction to topological superconductivity
  and Majorana fermions},
  \href{http://dx.doi.org/10.1088/0268-1242/27/12/124003}{Semicond. Sci.
  Technol. {\bf 27},  124003 } (2012).

\bibitem{elliott_colloquium_2015}
S.~R. Elliott and M. Franz, {\em Colloquium: Majorana fermions in nuclear, particle, and solid-state physics},
  \href{http://dx.doi.org/10.1103/RevModPhys.87.137}{Rev. Mod. Phys. {\bf 87},
  137 } (2015).

\bibitem{Beenakker2013Review}
C.~W.~J. Beenakker, {\em Search for Majorana fermions in superconductors},
  \href{http://dx.doi.org/10.1146/annurev-conmatphys-030212-184337}{Annu. Rev.
  Condens. Matter Phys. {\bf 4},  113 } (2013).

\bibitem{Lutchyn2010}
R.~M. Lutchyn, J.~D. Sau, and S. Das~Sarma, {\em Majorana fermions and a
  topological phase transition in semiconductor\-/superconductor
  heterostructures},
  \href{http://dx.doi.org/10.1103/PhysRevLett.105.077001}{Phys. Rev. Lett. {\bf
  105},  077001 } (2010).

\bibitem{Oreg2010}
Y. Oreg, G. Refael, and F. von Oppen, {\em Helical liquids and Majorana bound
  states in quantum wires},
  \href{http://dx.doi.org/10.1103/PhysRevLett.105.177002}{Phys. Rev. Lett. {\bf
  105},  177002 } (2010).

\bibitem{Gangadharaiah2011}
S. Gangadharaiah, B. Braunecker, P. Simon, and D. Loss, {\em Majorana edge
  states in interacting one-dimensional systems},
  \href{http://dx.doi.org/10.1103/PhysRevLett.107.036801}{Phys. Rev. Lett. {\bf
  107},  036801 } (2011).

\bibitem{Klinovaja2012}
J. Klinovaja, P. Stano, and D. Loss, {\em Transition from fractional to
  Majorana fermions in Rashba nanowires},
  \href{http://dx.doi.org/10.1103/PhysRevLett.109.236801}{Phys. Rev. Lett. {\bf
  109},  236801 } (2012).

\bibitem{Klinovaja2013}
J. Klinovaja, P. Stano, A. Yazdani, and D. Loss, {\em Topological
  superconductivity and Majorana fermions in RKKY systems},
  \href{http://dx.doi.org/10.1103/PhysRevLett.111.186805}{Phys. Rev. Lett. {\bf
  111},  186805 } (2013).

\bibitem{Jiang2011}
L. Jiang, D. Pekker, J. Alicea, G. Refael, Y. Oreg, and F. von Oppen, {\em
  Unconventional Josephson signatures of Majorana bound states},
  \href{http://dx.doi.org/10.1103/PhysRevLett.107.236401}{Phys. Rev. Lett. {\bf
  107},  236401 } (2011).

\bibitem{SanJose2012}
P. San-Jose, E. Prada, and R. Aguado, {\em ac Josephson effect in finite-length
  nanowire junctions with Majorana modes},
  \href{http://dx.doi.org/10.1103/PhysRevLett.108.257001}{Phys. Rev. Lett. {\bf
  108},  257001 } (2012).

\bibitem{Rokhinson2012}
L.~P. Rokhinson, X. Liu, and J.~K. Furdyna, {\em The fractional a.c. Josephson
  effect in a semiconductor-superconductor nanowire as a signature of Majorana
  particles}, \href{http://dx.doi.org/10.1038/nphys2429}{Nat. Phys. {\bf 8},
  795 } (2012).

\bibitem{Brunetti2013}
A. Brunetti, A. Zazunov, A. Kundu, and R. Egger, {\em Anomalous Josephson
  current, incipient time-reversal symmetry breaking, and Majorana bound states
  in interacting multilevel dots},
  \href{http://dx.doi.org/10.1103/PhysRevB.88.144515}{Phys. Rev. B {\bf 88},
  144515 } (2013).

\bibitem{Chang2013}
W. Chang, V.~E. Manucharyan, T.~S. Jespersen, J. Nyg\aa{}rd, and C.~M. Marcus,
  {\em Tunneling spectroscopy of quasiparticle bound states in a spinful
  Josephson junction},
  \href{http://dx.doi.org/10.1103/PhysRevLett.110.217005}{Phys. Rev. Lett. {\bf
  110},  217005 } (2013).

\bibitem{SanJose2013}
P. San-Jose, J. Cayao, E. Prada, and R. Aguado, {\em Multiple Andreev
  reflection and critical current in topological superconducting nanowire
  junctions}, New Journal of Physics  {\bf 15},  075019  (2013).

\bibitem{Dolcini2015}
F. Dolcini, M. Houzet, and J.~S. Meyer, {\em Topological Josephson
  ${\ensuremath{\phi}}_{0}$ junctions},
  \href{http://dx.doi.org/10.1103/PhysRevB.92.035428}{Phys. Rev. B {\bf 92},
  035428 } (2015).

\bibitem{Khanna2016}
U. Khanna, D.~K. Mukherjee, A. Kundu, and S. Rao, {\em Chiral nodes and
  oscillations in the Josephson current in Weyl semimetals},
  \href{http://dx.doi.org/10.1103/PhysRevB.93.121409}{Phys. Rev. B {\bf 93},
  121409 } (2016).

\bibitem{Peng2016}
Y. Peng, F. Pientka, E. Berg, Y. Oreg, and F. von Oppen, {\em Signatures of
  topological Josephson junctions},
  \href{http://dx.doi.org/10.1103/PhysRevB.94.085409}{Phys. Rev. B {\bf 94},
  085409 } (2016).

\bibitem{Hussein2016}
R. Hussein, L. Jaurigue, M. Governale, and A. Braggio, {\em Double quantum dot
  Cooper-pair splitter at finite couplings},
  \href{http://dx.doi.org/10.1103/PhysRevB.94.235134}{Phys. Rev. B {\bf 94},
  235134 } (2016).

\bibitem{Mellars2016}
E. Mellars and B. B\'eri, {\em Signatures of time-reversal-invariant
  topological superconductivity in the Josephson effect},
  \href{http://dx.doi.org/10.1103/PhysRevB.94.174508}{Phys. Rev. B {\bf 94},
  174508 } (2016).

\bibitem{Wiedenmann2016}
J. Wiedenmann, E. Bocquillon, R.~S. Deacon, S. Hartinger, O. Herrmann, T.~M.
  Klapwijk, L. Maier, C. Ames, C. Brune, C. Gould, A. Oiwa, K. Ishibashi, S.
  Tarucha, H. Buhmann, and L.~W. Molenkamp, {\em $4pi$-periodic Josephson
  supercurrent in HgTe-based topological Josephson junctions}, Nat. Commun.
  {\bf 7},  10303  (2016).

\bibitem{Jellinggaard2016}
A. Jellinggaard, K. Grove-Rasmussen, M.~H. Madsen, and J. Nyg\aa{}rd, {\em
  Tuning Yu-Shiba-Rusinov states in a quantum dot},
  \href{http://dx.doi.org/10.1103/PhysRevB.94.064520}{Phys. Rev. B {\bf 94},
  064520 } (2016).

\bibitem{Zyuzin2016}
A. Zyuzin, M. Alidoust, and D. Loss, {\em Josephson junction through a
  disordered topological insulator with helical magnetization},
  \href{http://dx.doi.org/10.1103/PhysRevB.93.214502}{Phys. Rev. B {\bf 93},
  214502 } (2016).

\bibitem{Deacon2017}
R.~S. Deacon, J. Wiedenmann, E. Bocquillon, F. Dom\'{\i}nguez, T.~M. Klapwijk,
  P. Leubner, C. Br{\"u}ne, E.~M. Hankiewicz, S. Tarucha, K. Ishibashi, H.
  Buhmann, and L.~W. Molenkamp, {\em Josephson radiation from gapless Andreev
  bound states in HgTe-based topological junctions},
  \href{http://dx.doi.org/10.1103/PhysRevX.7.021011}{Phys. Rev. X {\bf 7},
  021011 } (2017).

\bibitem{Virtanen2017}
P. Virtanen, F.~S. Bergeret, E. Strambini, F. Giazotto, and A. Braggio, {\em
  Majorana bound states in hybrid 2D Josephson junctions with ferromagnetic
  insulators}, \href{http://arxiv.org/abs/1712.01684}{arXiv:
  1712.01684   } (2017).

\bibitem{Mukherjee2017}
D.~K. Mukherjee, S. Rao, and A. Kundu, {\em Transport through Andreev bound
  states in a Weyl semimetal quantum dot},
  \href{http://dx.doi.org/10.1103/PhysRevB.96.161408}{Phys. Rev. B {\bf 96},
  161408 } (2017).

\bibitem{Khanna2017}
U. Khanna, S. Rao, and A. Kundu, {\em $0\text{\ensuremath{-}}\ensuremath{\pi}$
  transitions in a Josephson junction of an irradiated Weyl semimetal},
  \href{http://dx.doi.org/10.1103/PhysRevB.95.201115}{Phys. Rev. B {\bf 95},
  201115 } (2017).

\bibitem{Pientka2017}
F. Pientka, A. Keselman, E. Berg, A. Yacoby, A. Stern, and B.~I. Halperin, {\em
  Topological superconductivity in a planar Josephson junction},
  \href{http://dx.doi.org/10.1103/PhysRevX.7.021032}{Phys. Rev. X {\bf 7},
  021032 } (2017).

\bibitem{Hyang2017}
H. Huang, Q.-F. Liang, D.-X. Yao, and Z. Wang, {\em Majorana
  {\oe}{\"\i}0-junction in a disordered spin-orbit coupling nanowire with
  tilted magnetic field},
  \href{http://dx.doi.org/https://doi.org/10.1016/j.physc.2017.10.005}{Physica
  C: Superconductivity and its Applications {\bf 543},  22  } (2017).

\bibitem{Lee2017}
E.~J.~H. Lee, X. Jiang, R. \ifmmode~\check{Z}\else \v{Z}\fi{}itko, R. Aguado,
  C.~M. Lieber, and S. De~Franceschi, {\em Scaling of subgap excitations in a
  superconductor-semiconductor nanowire quantum dot},
  \href{http://dx.doi.org/10.1103/PhysRevB.95.180502}{Phys. Rev. B {\bf 95},
  180502 } (2017).

\bibitem{Cayao2017}
J. Cayao, P. San-Jose, A.~M. Black-Schaffer, R. Aguado, and E. Prada, {\em
  Majorana splitting from critical currents in Josephson junctions},
  \href{http://dx.doi.org/10.1103/PhysRevB.96.205425}{Phys. Rev. B {\bf 96},
  205425 } (2017).

\bibitem{Dominguez2017}
F. Dom\'{\i}nguez, O. Kashuba, E. Bocquillon, J. Wiedenmann, R.~S. Deacon,
  T.~M. Klapwijk, G. Platero, L.~W. Molenkamp, B. Trauzettel, and E.~M.
  Hankiewicz, {\em Josephson junction dynamics in the presence of
  $2\ensuremath{\pi}$- and $4\ensuremath{\pi}$-periodic supercurrents},
  \href{http://dx.doi.org/10.1103/PhysRevB.95.195430}{Phys. Rev. B {\bf 95},
  195430 } (2017).

\bibitem{Cayao2018}
J. Cayao, A.~M. Black-Schaffer, E. Prada, and R. Aguado, {\em Andreev spectrum
  and supercurrents in nanowire-based SNS junctions containing Majorana bound
  states}, \href{http://dx.doi.org/10.3762/bjnano.9.127}{Beilstein Journal of
  Nanotechnology {\bf 9},  1339 } (2018).

\bibitem{Marra2016}
P. Marra, R. Citro, and A. Braggio, {\em Signatures of topological phase
  transitions in Josephson current-phase discontinuities},
  \href{http://dx.doi.org/10.1103/PhysRevB.93.220507}{Phys. Rev. B {\bf 93},
  220507 } (2016).

\bibitem{Loss1998}
D. Loss and D.~P. DiVincenzo, {\em Quantum computation with quantum dots},
  \href{http://dx.doi.org/10.1103/PhysRevA.57.120}{Phys. Rev. A {\bf 57},  120
  } (1998).

\bibitem{Choi2000}
M.-S. Choi, C. Bruder, and D. Loss, {\em Spin-dependent Josephson current
  through double quantum dots and measurement of entangled electron states},
  \href{http://dx.doi.org/10.1103/PhysRevB.62.13569}{Phys. Rev. B {\bf 62},
  13569 } (2000).

\bibitem{Kouwenhoven2001}
L.~P. Kouwenhoven, D.~G. Austing, and S. Tarucha, {\em Few-electron quantum
  dots}, Reports on Progress in Physics  {\bf 64},  701  (2001).

\bibitem{Wiel2001}
W.~G. van~der Wiel, T. Fujisawa, S. Tarucha, and L.~P. Kouwenhoven, {\em A
  double quantum dot as an artificial two-level system}, Japanese Journal of
  Applied Physics  {\bf 40},  2100  (2001).

\bibitem{Tarasinski2015}
B. Tarasinski, D. Chevallier, J.~A. Hutasoit, B. Baxevanis, and C.~W.~J.
  Beenakker, {\em Quench dynamics of fermion-parity switches in a Josephson
  junction}, \href{http://dx.doi.org/10.1103/PhysRevB.92.144306}{Phys. Rev. B
  {\bf 92},  144306 } (2015).

\bibitem{Beenakker2013}
C.~W.~J. Beenakker, J.~M. Edge, J.~P. Dahlhaus, D.~I. Pikulin, S. Mi, and M.
  Wimmer, {\em Wigner-Poisson statistics of topological transitions in a
  Josephson junction},
  \href{http://dx.doi.org/10.1103/PhysRevLett.111.037001}{Phys. Rev. Lett. {\bf
  111},  037001 } (2013).

\bibitem{Stanescu2013}
T.~D. Stanescu, R.~M. Lutchyn, and S. Das~Sarma, {\em Dimensional crossover in
  spin-orbit-coupled semiconductor nanowires with induced superconducting
  pairing}, \href{http://dx.doi.org/10.1103/PhysRevB.87.094518}{Phys. Rev. B
  {\bf 87},  094518 } (2013).

\bibitem{Lee2013}
E.~J.~H. Lee, X. Jiang, M. Houzet, R. Aguado, C.~M. Lieber, and S.
  De~Franceschi, {\em Spin-resolved Andreev levels and parity crossings in
  hybrid superconductor--semiconductor nanostructures}, Nature Nanotechnology
  {\bf 9},  79 EP   (2013).

\bibitem{Yokoyama2013}
T. Yokoyama, M. Eto, and Y.~V. Nazarov, {\em Josephson current through
  semiconductor nanowire with spin-orbit interaction in magnetic field},
  \href{http://dx.doi.org/10.7566/JPSJ.82.054703}{J. Phys. Soc. Jpn. {\bf 82},
  054703 } (2013).

\bibitem{Yokoyama2014}
T. Yokoyama, M. Eto, and Y.~V. Nazarov, {\em Anomalous Josephson effect induced
  by spin\-/orbit interaction and Zeeman effect in semiconductor nanowires},
  \href{http://dx.doi.org/10.1103/PhysRevB.89.195407}{Phys. Rev. B {\bf 89},
  195407 } (2014).

\bibitem{Klees2017}
R.~L. Klees, G. Rastelli, and W. Belzig, {\em Nonequilibrium Andreev bound
  states population in short superconducting junctions coupled to a resonator},
  \href{http://dx.doi.org/10.1103/PhysRevB.96.144510}{Phys. Rev. B {\bf 96},
  144510 } (2017).

\bibitem{Yantao2014}
Y. Li, A. Kundu, F. Zhong, and B. Seradjeh, {\em Tunable Floquet Majorana
  fermions in driven coupled quantum dots},
  \href{http://dx.doi.org/10.1103/PhysRevB.90.121401}{Phys. Rev. B {\bf 90},
  121401 } (2014).

\bibitem{Benito2015}
M. Benito and G. Platero, {\em Floquet Majorana fermions in superconducting
  quantum dots}, \href{http://dx.doi.org/10.1016/j.physe.2015.08.030}{Physica
  E: Low-dimensional Systems and Nanostructures {\bf 74},  608  } (2015).

\bibitem{Meng2009}
T. Meng, S. Florens, and P. Simon, {\em Self-consistent description of Andreev
  bound states in Josephson quantum dot devices},
  \href{http://dx.doi.org/10.1103/PhysRevB.79.224521}{Phys. Rev. B {\bf 79},
  224521 } (2009).

\bibitem{Wentzell2016}
N. Wentzell, S. Florens, T. Meng, V. Meden, and S. Andergassen, {\em
  Magnetoelectric spectroscopy of Andreev bound states in Josephson quantum
  dots}, \href{http://dx.doi.org/10.1103/PhysRevB.94.085151}{Phys. Rev. B {\bf
  94},  085151 } (2016).

\bibitem{Braggio2011}
A. Braggio, M. Governale, M.~G. Pala, and J. K{\"o}nig, {\em Superconducting
  proximity effect in interacting quantum dots revealed by shot noise},
  \href{http://dx.doi.org/https://doi.org/10.1016/j.ssc.2010.10.043}{Solid
  State Communications {\bf 151},  155  } (2011).

\bibitem{Droste2012}
S. Droste, S. Andergassen, and J. Splettstoesser, {\em Josephson current
  through interacting double quantum dots with spin-orbit coupling}, Journal of
  Physics: Condensed Matter  {\bf 24},  415301  (2012).

\bibitem{Hussein2017}
R. Hussein, A. Braggio, and M. Governale, {\em Entanglement-symmetry control in
  a quantum-dot Cooper-pair splitter},
  \href{http://dx.doi.org/10.1002/pssb.201600603}{Physica Status Solidi (B)
  {\bf 254},  1600603 } (2017).

\bibitem{Nazarov2009Page106}
Y. Nazarov and Y. Blanter, {\em Quantum Transport: Introduction to Nanoscience}
  (Cambridge University Press, 2009), p.\ 106.

\bibitem{Loring2015}
T.~A. Loring, {\em K-theory and pseudospectra for topological insulators},
  \href{http://dx.doi.org/10.1016/j.aop.2015.02.031}{Ann. Phys. {\bf 356},  383
   } (2015).

\bibitem{Golubov2004}
A.~A. Golubov, M.~Y. Kupriyanov, and E. Il'ichev, {\em The current-phase
  relation in Josephson junctions},
  \href{http://dx.doi.org/10.1103/RevModPhys.76.411}{Rev. Mod. Phys. {\bf 76},
  411 } (2004).

\bibitem{Tiira2017}
J. Tiira, E. Strambini, M. Amado, S. Roddaro, P. San-Jose, R. Aguado, F.~S.
  Bergeret, D. Ercolani, L. Sorba, and F. Giazotto, {\em Magnetically-driven
  colossal supercurrent enhancement in InAs nanowire Josephson junctions},
  \href{http://dx.doi.org/10.1038/ncomms14984}{Nature Communications {\bf 8},
  14984 } (2017).

\bibitem{Frolov2004}
S.~M. Frolov, D.~J. Van~Harlingen, V.~A. Oboznov, V.~V. Bolginov, and V.~V.
  Ryazanov, {\em Measurement of the current-phase relation of
  superconductor/ferromagnet/superconductor $\ensuremath{\pi}$ Josephson
  junctions}, \href{http://dx.doi.org/10.1103/PhysRevB.70.144505}{Phys. Rev. B
  {\bf 70},  144505 } (2004).

\bibitem{Sochnikov2013}
I. Sochnikov, A.~J. Bestwick, J.~R. Williams, T.~M. Lippman, I.~R. Fisher, D.
  Goldhaber-Gordon, J.~R. Kirtley, and K.~A. Moler, {\em Direct measurement of
  current-phase relations in superconductor/topological
  insulator/superconductor junctions},
  \href{http://dx.doi.org/10.1021/nl400997k}{Nano Letters {\bf 13},  3086 }
  (2013).

\bibitem{Szombati2016}
D.~B. Szombati, S. Nadj-Perge, D. Car, S.~R. Plissard, E.~P. A.~M. Bakkers, and
  L.~P. Kouwenhoven, {\em Josephson $\phi_0$-junction in nanowire quantum
  dots}, Nature Physics  {\bf 12},  568  (2016).

\bibitem{Ozyuzer2007}
L. Ozyuzer, A.~E. Koshelev, C. Kurter, N. Gopalsami, Q. Li, M. Tachiki, K.
  Kadowaki, T. Yamamoto, H. Minami, H. Yamaguchi, T. Tachiki, K.~E. Gray, W.-K.
  Kwok, and U. Welp, {\em Emission of coherent THz radiation from
  superconductors}, \href{http://dx.doi.org/10.1126/science.1149802}{Science
  {\bf 318},  1291 } (2007).

\bibitem{Solinas2015}
P. Solinas, S. Gasparinetti, D. Golubev, and F. Giazotto, {\em A Josephson
  radiation comb generator}, Scientific Reports  {\bf 5},  12260  (2015).

\end{thebibliography}

\end{document}